\documentclass{aastex631}%

\usepackage{amsmath}
\usepackage{amssymb}
\usepackage{amsfonts}
\usepackage{graphicx}
\usepackage{hyperref}%
\providecommand{\U}[1]{\protect\rule{.1in}{.1in}}
\begin{document}
\title{Catalogue of wide binary, trinary and quaternary candidates from the Gaia data
release 2\\ (region $\left\vert b\right\vert >25\,\deg$)}
\author{Petr Zavada}
\author{Karel P\'{\i}\v{s}ka}
\affiliation{Institute of Physics of the Czech Academy of Sciences, Na Slovance 2, 182 21
Prague 8, Czech Republic}
\accepted{by the Astronomical Journal: October, 2021}

\begin{abstract}
The occurrence of multiple stars, dominantly binaries, is studied using the
\textit{Gaia}-ESA DR2 catalogue. We apply the optimized statistical method
that we previously developed for the analysis of 2D patterns. The field of
stars is divided into a mosaic of small pieces, which represent a statistical
set for analysis. Specifically, data input is represented by a grid of circles
(events) with radius $0.02\,\deg$ covering the sky in the field of galactic
latitude $\left\vert b\right\vert >25\,\deg$. The criteria for selecting
candidates for multiple stars are based on two parameters: angular separation
and collinearity of proper motion. Radial separation, due to limited accuracy,
is used only as a weaker supplementary constraint. Due attention is paid to
the accurate calculation of the background, which is a necessary input for
evaluating the quality of the candidates. Our selection algorithm generates
the catalogue of candidates: $900,842$ binaries, $5,282$ trinaries and $30$ quaternaries.

\end{abstract}

\section{ Introduction}

It is believed that statistics of binaries and multiple stars can provide
deeper insights into the formation and evolution of galaxies. Wide binaries
may serve as a sensitive probe of the Galactic gravitational potential. Some
recent studies suggest that very wide binaries may provide data on the
presence of dark matter in the Galaxy (\cite{pena,hern,pitt}). At present, an
extremely rich source of data on stars of our Galaxy is provided by the
\textit{Gaia}-ESA mission and collected in the catalogue DR2 (\cite{gaia2}).
The release of the next and more expanded full version (DR3) is expected in
the first half of 2022. Several authors have recently studied various aspects
of binaries using the \textit{Gaia} DR1 (\cite{oelkers,semyeong}), DR2
(\cite{ziegler,badry18,godo,esteban,gonz,sapo,hart,tian}) and EDR3 (Early Data
Release 3) (\cite{badry,gaia3}) catalogues. Other important articles examining
wide binaries before Gaia include, for example, \cite{cabalero} and
\cite{close}.

In our previous study (\cite{AApzkp,AJpzkp}) we have developed and applied
statistical methods for the identification of binaries and multiple stars with
an accurate estimate of the background. The resulting catalogue is based on
the 3D event analysis and contains about $8\times10^{4}$ binary candidates at
a distance of up to $\lesssim$340\thinspace pc. This catalogue involves the
binary stars whose total separation does not exceed the event sphere diameter
that was $4$\thinspace pc. Because determining the radial separation $Z_{ij}$
can have a significant error, we have been selecting the binary stars in the
events based on the distribution of the projected distances $\Delta_{ij}$, the
error of which is only slightly affected by the parallax error. At the same
time, the error of parallax and radial distance causes that the constraint on
radial separation $Z_{ij}\leq$ $4$\thinspace pc can exclude a large number of
true binary stars. In the present study, we solve this drawback of 3D events
by a more general procedure. The procedure combines the analysis of angular
separations $d_{ij}$ of sources inside 2D events with the information on the
radial separation $Z_{ij}$. Simultaneously, we apply the condition of
approximate collinearity of proper motion here as before. The result of the
procedure is the selection of candidates for the binary with the defined
probability that it is the true binary and not a random background.

Our 2D analysis is described in Sec.\ref{SEC2} and consists of several parts.
The basic notions, which our method deals with are defined in section
\ref{meth}. Input data from the sector of the \textit{Gaia} catalogue are
defined in section \ref{inda}. In the next section \ref{sebi} we analyze the
probability of the binary depending on the angular separation and collinearity
of proper motion of its components. This analysis sets the rules for the
selection of binary candidates for our new catalogue. Using parallax, we can
define the distance of binary and the projection of the absolute separation,
which is more important for physics than just the angular separation. In
section \ref{pasp} we show that distributions of the projection and
correlations with proper motion allow us to obtain\ information about binary
orbit. In section \ref{witq} we show the results of our analysis on the
occurrence of trinaries and quaternaries. \ Candidates with a high degree of
reliability are listed in the electronic catalogue. Its structure and content
are described in Sec.\ref{SEC3}. The comparison with the other catalogues
(\cite{badry,hart,esteban,AJpzkp}) is done in Sec.\ref{SEC4}. The last
Sec.\ref{SEC5} is devoted to the summary and concluding remarks.

\section{Statistical analysis of 2D patterns}

\label{SEC2}

\subsection{Methodology}

\label{meth}The method of 2D analysis is described in detail in our previous
paper \cite{AJpzkp}, so here we repeat only basic notions.

\begin{figure}[t]
\centering\includegraphics[width=80mm]{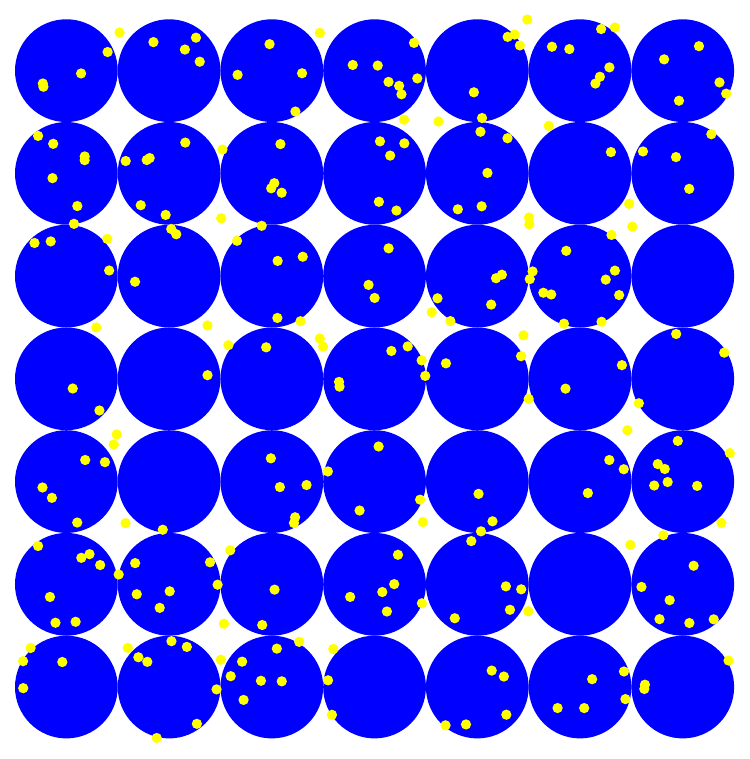} \caption{Grid of 2D event
circles with uniform distribution of the stars.}%
\label{ff0}%
\end{figure}

The data for analysis are represented by the grid of circles with patterns of
stars covering a defined region of sky (Fig.\ref{ff0}). The input data for generating the grid are given in the galactic
reference frame. So, the position $\mathbf{L}$ of a source is defined by
spherical coordinates $L,l$ and $b$ (distance from the Sun, galactic longitude
and latitude):%
\begin{gather}
\mathbf{L}=L\mathbf{n;\qquad n}=(\cos b\cos l,\cos b\sin l,\sin b),\label{sa1}%
\\
-\frac{\pi}{2}\leq b\leq\frac{\pi}{2},\qquad-\pi<l\leq\pi.\nonumber
\end{gather}
In the centre of each circle we define local orthonormal frame defined by the
basis:
\begin{align}
\mathbf{k}_{r}  &  =\mathbf{n}_{0}=(\cos b_{0}\cos l_{0},\cos b_{0}\sin
l_{0},\sin b_{0}),\label{sa3}\\
\mathbf{k}_{l}  &  =(-\sin l_{0},\cos l_{0},0),\nonumber\\
\mathbf{k}_{b}  &  =(-\sin b_{0}\cos l_{0},-\sin b_{0}\sin l_{0},\cos
b_{0}),\nonumber
\end{align}
where $\mathbf{k}_{r}=\mathbf{n}_{0}\left(  b_{0},l_{0}\right)  $ defines
angular position of the circle centre. Unit vector $\mathbf{k}_{l}$ is
perpendicular to $\mathbf{k}_{r}$ and has direction of increasing $l$. Unit
vector $\mathbf{k}_{b}$ is defined as $\mathbf{k}_{b}=\mathbf{k}_{r}%
\times\mathbf{k}_{l}$ and has direction of increasing $b$. Vector
$\mathbf{k}_{r}$ has radial direction, perpendicular vectors $\mathbf{k}_{b}$
and $\mathbf{k}_{l}$ \ lies in the transverse plain. The stars inside the
circle of small radius $\rho_{2}$ satisfy
\begin{equation}
\left\vert \mathbf{n}_{i}-\mathbf{n}_{0}\right\vert \leq\rho_{2},\qquad
i=1,...M. \label{sa2}%
\end{equation}
or%
\begin{equation}
\{x_{i},y_{i}\};\quad x_{i}^{2}+y_{i}^{2}\leq\rho_{2}^{2},\quad i=1,...M,
\label{sa5}%
\end{equation}
where $\{x_{i},y_{i}\}$ are local rectangular coordinates defined by the basis
(\ref{sa3}):
\begin{equation}
x_{i}=\mathbf{n}_{i}^{\prime}.\mathbf{k}_{l},\qquad y_{i}=\mathbf{n}%
_{i}^{\prime}.\mathbf{k}_{b};\qquad\mathbf{n}_{i}^{\prime}=\mathbf{n}%
_{i}-\mathbf{n}_{0}. \label{sa4}%
\end{equation}
The circles in the grid are, in fact, spherical caps. However, their radius
$\rho_{2}$ will be so small $(\simeq3.5\times10^{-4}$rad$)$, that the caps can
be reliably considered as flat circles.

A set of stars defined by (\ref{sa5}) is called the \textit{event}. We define
the pair separations%
\begin{equation}
x_{ij}=\left\vert x_{j}-x_{i}\right\vert ;\qquad y_{ij}=\left\vert y_{j}%
-y_{i}\right\vert ;\qquad d_{ij}=\sqrt{x_{ij}^{2}+y_{ij}^{2}};\qquad
i,j=1,2,...M. \label{SA21}%
\end{equation}
It is useful to define the scaled separation%
\begin{equation}
\hat{\xi}=\frac{d_{ij}}{2\rho_{2}};\qquad0\leq\hat{\xi}\leq1, \label{SA22}%
\end{equation}
where $\rho_{2}$ is the angular radius of the events. Distribution of
separations of randomly distributed sources inside the circle \textit{does not
depend} on $M$ and is given by relation%
\begin{equation}
q\left(  \hat{\xi}\right)  =\frac{16\hat{\xi}}{\pi}\left(  \arccos\hat{\xi
}-\hat{\xi}\sqrt{1-\hat{\xi}^{2}}\right)  , \label{SA23}%
\end{equation}
which was proved in (\cite{AJpzkp}). This curve is important for subtraction
of random background.

Before practical use, we add a few general remarks:

i) Of course, the grid in Fig.\ref{ff0}\ can be rectangular only locally. In
fact, we work with circles with the same radius aligned only in rows with a
constant galactic latitude.

ii) The radius $\rho_{2}$ must be set to be significantly larger than a
typical separation of true binary. At the same time, it must be so small that
the distribution of stars within the event can be considered random and
uniform. In Sec. \ref{sebi} these conditions will be explained in more detail.
We will also explain that the obtained results are not sensitive to the exact
setting of $\rho_{2}$.

iii) Events with too high multiplicity $M$, in which various higher dense
structures may dominate, are excluded from processing.

iv) The circular shape of the events is chosen due to a simple but accurate
formula (\ref{SA23}) for calculating a random background. Another shape would
lead to a more complex formula depending on other shape parameters (triangle,
square, orientation ...). On the other hand, the grid of circle events only
partially covers the sky. The complete set of binaries is obtained in
Sec.\ref{SEC3} from several shifted grids.

\subsection{Input data}

\label{inda}
\begin{figure}[t]
\centering\includegraphics[width=80mm]{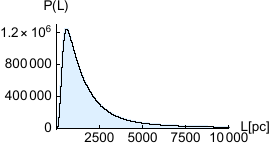}\caption{Distribution of
distances of stars of the region $\mathbf{R}$. Binning: 30pc.}%
\label{ff1}%
\end{figure}

The parameters of the events we work with are listed in
Tab.\ref{Tab1}. \begin{table}[ptb]
\begin{center}%
\begin{tabular}
[c]{cccccccc}
& 2D region: $l\times b[\deg^{2}]$ & $\rho_{2}[\mathrm{as}]$ & $\left\langle
L\right\rangle \left[  \text{pc}\right]  $ & $\left\langle M\right\rangle $ &
$N_{e}$ & $N_{s}$ & $N_{tot}$\\\hline\hline
R$_{1}$ & $\left\langle -180,180\right\rangle \times\left\langle \pm
45,\pm90\right\rangle $ & $72$ & $1,807$ & $3.86$ & $5,887,737$ & $22,083,670$
& $30,643,238$\\\hline
R$_{2}$ & $\left\langle -180,180\right\rangle \times\left\langle \pm
25,\pm45\right\rangle $ & $72$ & $2,134$ & $7.86$ & $6,985,043$ & $53,043,629$
& $80,653,496$%
\end{tabular}
\end{center}
\caption{Analyzed regions \textbf{R}$_{1,2}$ in the DR2 catalogue, where
$\rho_{2}$ is the angular radius of the events, $\left\langle L\right\rangle
,\left\langle M\right\rangle $ are average distance and event multiplicity,
$N_{e}$ is the total number of events. Only sources with positive parallax and
in distance $<$ $15,000\,$pc are taken into account and only events $2\leq
M\leq25$ are accepted for present analysis$.$ $N_{s}$ is number of sources
after these cuts, $N_{tot}$ is total number of sources with positive parallax
in \textbf{R}$_{1,2}$.}%
\label{Tab1}%
\end{table}By symbol $\mathbf{R}$ we denote the union of both regions in
table:%
\begin{equation}
\mathbf{R=R}_{1}\cup\mathbf{R}_{2}. \label{SA01}%
\end{equation}
In Fig.\ref{ff1} we have shown distribution of the distances of the stars in the
table from the Sun (origin of the galactic reference frame). If the parallax
$p$ is given in angular units \thinspace\textbf{as} ($=1^{\prime\prime}$),
then the corresponding distance is%
\begin{equation}
L[\mathrm{pc}]=\frac{1}{p[\mathrm{as}]}. \label{SA0z}%
\end{equation}
The figure demonstrates the scale of distances measured by \textit{Gaia}. This
histogram also suggests the convention valid for all histograms P($x$) in this
paper: ordinate represents the number of entries of variable $x$ of given
value into the bins of the width specified in the figure caption.

\begin{figure}[t]
\centering\includegraphics[width=80mm]{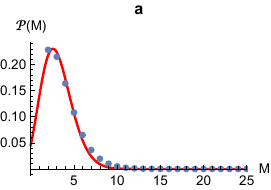}\includegraphics[width=80mm]{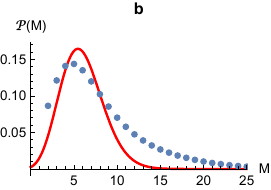}\caption{Distribution
of multiplicities of events from regions \textbf{R}$_{1}$(panel\textit{ a})
and \textbf{R}$_{2}$ (panel\textit{ b}). Red curves represent Poisson
distributions, where discrete values $M!$ are interpolated by gamma function
$\Gamma(M+1)$.}%
\label{ff2}%
\end{figure}

Stars in the sky are divided into the circle regions (events) of angular
radius $\rho_{2}$ and in Fig.\ref{ff2} we show distribution of the star multiplicities in these events.
The distribution of stars in region $\mathbf{R}_{1}$ is roughly homogenous, so
the multiplicity distribution has a nearly Poisson shape%
\begin{equation}
\mathcal{P}(M)=\frac{\lambda^{M}}{M!}\exp(-\lambda);\qquad\lambda=\left\langle
M\right\rangle . \label{SA0}%
\end{equation}
Since region $\mathbf{R}_{2}$ shows greater density fluctuations resulting in
a broader distribution then Poisson, we fail to make its fit. This region is
dominated by an inhomogeneous stellar density due to patchy inter-stellar
extinction, the presence of open clusters, and the imprint of the Galactic
spiral pattern.

The proper motion of the stars in DR2 is defined by two angular velocities%
\begin{equation}
\mu_{\alpha}^{\ast}(\equiv\mu_{\alpha}\cos\delta),\qquad\mu_{\delta}
\label{SABE13}%
\end{equation}
in directions of the right ascension and declination in the ICRS. So the
corresponding transverse 2D velocity $\mathbf{U}$ is given as%
\begin{equation}
\mathbf{U=}L\mathbf{u},\qquad\mathbf{u=(}\mu_{\alpha}^{\ast},\mu_{\delta
}),\qquad u\mathbf{=}\left\vert \mathbf{u}\right\vert . \label{SABE14}%
\end{equation}
For the pair of stars we can define the angle between both transverse
velocities:%
\begin{equation}
\alpha_{ij}=\arccos\frac{\mathbf{u}_{i}\cdot\mathbf{u}_{j}}{u_{i}u_{j}}.
\label{SABE15}%
\end{equation}

\begin{figure}[t]
\centering\includegraphics[width=80mm]{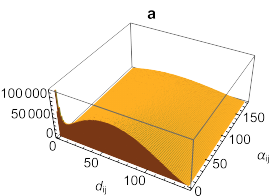}\includegraphics[width=80mm]{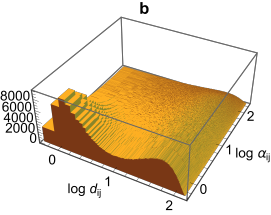}
\includegraphics[width=80mm]{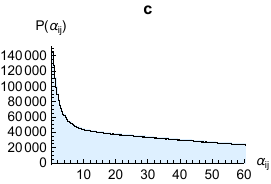}\includegraphics[width=80mm]{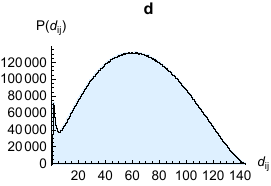}\caption{Signature
of binaries in region $\mathbf{R}$: distribution $P(d_{ij},\alpha_{ij})$
(panels \textit{a,b - }linear and logarithmic scales), distribution
$P(\alpha_{ij})$ for $d_{ij}\leq15\,\mathrm{as}$ (panel \textit{c}) and
distribution $P(d_{ij})$ for $\alpha_{ij}\leq15\deg$ (panel \textit{d}).
Units: $d_{ij}[\mathrm{as}],\alpha_{ij}[\deg].$ Binning: $1.44$as$\times
1.8\deg,0.36as\times0.45\deg,0.36\deg,0.288$as.}%
\label{ff3}%
\end{figure}

Our selection of binaries is based on Fig.\ref{ff3} obtained from all events in the table. The peak in the domain of
small separations $d_{ij}$ and small angles $\alpha_{ij}$ represents a clear
signature of binaries (panel \textit{a}). \ The domain of the peak can be
approximately defined by the conditions%
\begin{equation}
d_{ij}\leq15\,\mathrm{as},\qquad\alpha_{ij}\leq15\deg. \label{SA0a}%
\end{equation}
The second condition was applied also in our former 3D analysis. Panel
\textit{b} shows distribution $P(\alpha_{ij})$ in the band $d_{ij}%
\leq15\,\mathrm{as}$, similarly in panel \textit{c} we have distribution
$P(d_{ij})$ for $\alpha_{ij}\leq15\deg$. Recall that the stars separated by
$d_{ij}\lesssim0.5\,\mathrm{as}$ are missing due to current resolution limit
in the DR2 data set as stated in (\cite{arenou2}).

\begin{figure}[t]
\centering\includegraphics[width=80mm]{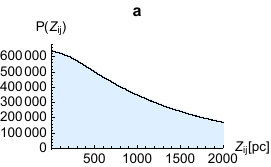}\includegraphics[width=80mm]{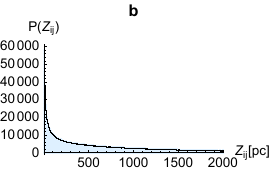}\caption{Radial
separations for pairs of the region $\mathbf{R}$ outside window (\ref{SA0a})
(panel \textit{a}) and inside the window (panel \textit{b}). Binning: $4$pc.}%
\label{ff4}%
\end{figure}

\begin{figure}[t]
\centering\includegraphics[width=60mm]{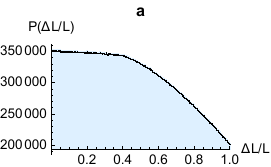}\includegraphics[width=60mm]{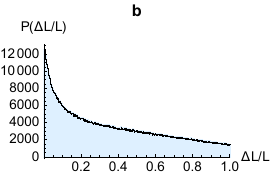}\includegraphics[width=60mm]{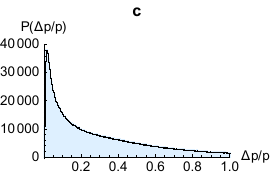}\caption{Comparison
of distributions of relative errors: $\Delta L/L$ outside (panel \textit{a}),
inside (panel \textit{b}) the window of binaries (\ref{SA0a}) and $\ \Delta
p/p$ (panel \textit{c}) for sources inside the window. Data are from the
region $\mathbf{R}$. Binning: 0.002.}%
\label{ff5LP}%
\end{figure}

In Fig.\ref{ff4} we have shown distribution of radial separations $Z_{ij}$ defined
as%
\begin{equation}
Z_{ij}=\left\vert L_{i}-L_{j}\right\vert \label{SA0b}%
\end{equation}
and obtained with the use of parallax (\ref{SA0z}). Panel \textit{a(b)} shows
distribution of pairs outside (inside) the window (\ref{SA0a}). The sharp peak
at small radial separations in the second panel is expected for binaries, the
tail with greater separations corresponds partly to the background pairs and
partly to the true binaries having large errors of parallaxes. Alternative
distributions $P(\Delta L/L)$, where
\begin{equation}
\Delta L\approx Z_{ij},\qquad L=\frac{L_{i}+L_{j}}{2}, \label{SA0c}%
\end{equation}
are shown in Fig.\ref{ff5LP}\textit{a,b}. Relation (\ref{SA0z}) implies%
\begin{equation}
\frac{\Delta L}{L}\approx\frac{\Delta p}{p}. \label{SA0d}%
\end{equation}
In Fig.\ref{ff5LP}\textit{c} we show distribution $P(\Delta p/p)$ obtained
directly from the \textit{Gaia} data, where the parallax of the source is
accompanied by its estimated error. Our $\Delta L,L$ related to binaries are
also obtained from parallaxes, however, our estimate of relative error of the
parallax is independent and can serve as a cross-check. We observe a
noticeable similarity between the distributions \textit{b,c}. A smaller
difference can arise because the window of binaries also contains a background
with random radial separations.

\subsection{Selection of binaries and calculation of background}

\label{sebi}In the next we will work with the numbers:

$n_{2}-$ number of (true) binaries

$b_{2}-$ number of background pairs

$N_{2}-$ total number of pairs:
\begin{equation}
N_{2}=n_{2}+b_{2}. \label{SA34}%
\end{equation}
These numbers are related to given sample of pairs defined by the
corresponding cuts. One cannot decide if the pair is a binary or background,
but we can find out the probability that the pair is binary $\left(
n_{2}/N_{2}\right)  $ or background $\left(  b_{2}/N_{2}\right)  $. Obviously,
in the domain of the peak in Fig.\ref{ff3}\textit{a }the probability of
binaries $n_{2}/N_{2}$ is high.

Radial separation $Z_{ij}$ can be used to further increase the probability of
binaries in the peak, but due to low accuracy, the cut must be set
judiciously. Too strict cut generates a cleaner sample of binaries (higher
ratio $n_{2}/N_{2}$) but more binaries are excluded. And conversely, too soft
cut preserves more binaries, but at the price of the higher background and
lower ratio $n_{2}/N_{2}$.

The area $d\times\alpha$ of Fig.\ref{ff3}\textit{a} can be divided into four
the windows \textbf{A,B,C} and \textbf{D} (like in Fig.\ref{ff7}):
\begin{equation}
\mathbf{A}=\left\langle 0,d_{c}\right\rangle \times\left\langle 0,\alpha
_{c}\right\rangle ;\qquad d_{c}=15\,\mathrm{as},\qquad\alpha_{c}=15\deg,
\label{SA4}%
\end{equation}
which is the domain of the peak (\ref{SA0a}) with the high population of
binaries. The remaining windows are%

\begin{align}
\mathbf{B}  &  =\left\langle d_{c},d_{\max}\right\rangle \times\left\langle
0,\alpha_{c}\right\rangle ,\label{SA4a}\\
\mathbf{C}  &  =\left\langle d_{c},d_{\max}\right\rangle \times\left\langle
\alpha_{c},180\right\rangle ,\nonumber\\
\mathbf{D}  &  =\left\langle 0,d_{c}\right\rangle \times\left\langle
\alpha_{c},180\right\rangle .\nonumber
\end{align}

\begin{figure}[t]
\centering\includegraphics[width=60mm]{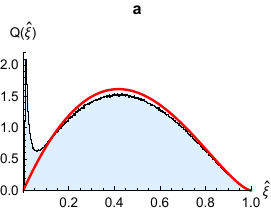}\includegraphics[width=60mm]{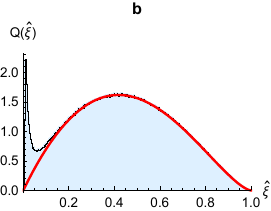}\includegraphics[width=60mm]{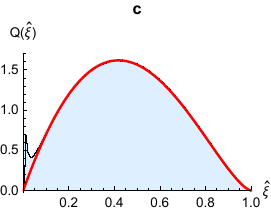}\caption{Scaled
separation $\hat{\xi}$: the peak of binaries and random background
(\textit{red curve}). Region \textbf{R}$_{1}$: equal normalization of data and
background (panel\textit{\ a}), data and renormalized background
(panel\ \textit{b}). Region \textbf{R}$_{2}$: data and renormalized background
(panel\ \textit{c}).}%
\label{ff5}%
\end{figure}

\begin{figure}[t]
\centering\includegraphics[width=160mm]{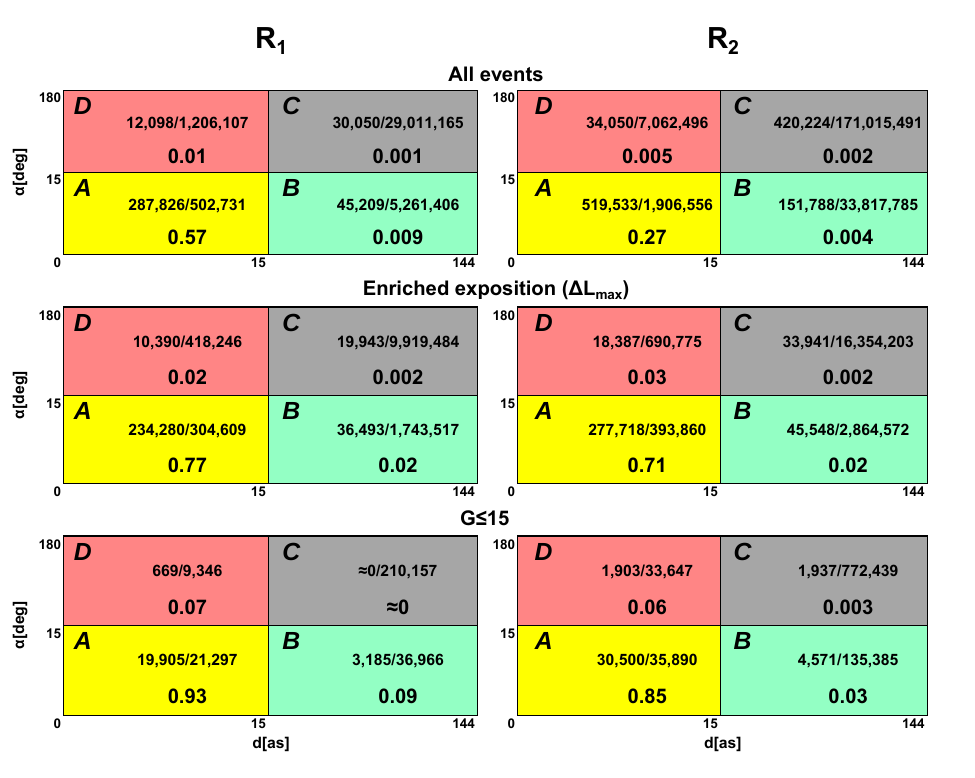}\caption{Ratio $n_{2}/N_{2}$
of the number of binaries and the total number of pairs in different
$d\times\alpha$\ windows in the regions \textbf{R}$_{1}$ and \textbf{R}$_{2}$
under different conditions. The result of this ratio $\left(  \beta\right)  $
is in the second row of the respective window.}%
\label{ff7}%
\end{figure}

\noindent\qquad Distribution $P(d_{ij})$ \ can be represented by its
normalized form%
\begin{equation}
Q(\hat{\xi})=\frac{1}{N_{2}}P(d_{ij});\quad\hat{\xi}=\frac{d_{ij}}{2\rho_{2}}.
\label{SA5}%
\end{equation}
The random background is described by the normalized function $q(\hat{\xi})~$
defined by Eq.(\ref{SA23}). Both distributions are shown in Fig.\ref{ff5}%
\textit{a }for region $\mathbf{R}_{1}$. Due to equal normalization, an excess in the peak of binaries in
distribution $P$ is compensated by the lack of pairs in the region of
background. Assuming that for $d_{ij}\geq2/3d_{\max}$ or $\hat{\xi}\geq2/3$
distribution $P$ involves only background, we renormalize $q$ correspondingly:%
\begin{equation}
q(\hat{\xi})\rightarrow q^{r}(\hat{\xi})=\gamma q(\hat{\xi});\qquad
\gamma=\frac{\int_{\hat{\xi}\geq2/3}Q(\hat{\xi})d\hat{\xi}}{\int_{\hat{\xi
}\geq2/3}q(\hat{\xi})d\hat{\xi}}. \label{SA7}%
\end{equation}
In panel \textit{b} we have shown distribution $Q(\hat{\xi})$ together with
the renormalized background $q^{r}(\hat{\xi})$. The same distributions for
region $\mathbf{R}_{2}$ is shown in panel \textit{c}. Obviously, the
distribution of binaries is given by their difference%
\begin{equation}
P_{bin}(\hat{\xi})=N_{2}\left(  Q(\hat{\xi})-q^{r}(\hat{\xi})\right)  .
\label{SA8}%
\end{equation}
It is now clear from these figures why the diameter of events $2\rho_{2}$ is
chosen much greater than typical separation of the binaries $d_{bin}$ \ (the
parameter $\hat{\xi}_{bin}=d_{bin}/2\rho_{2}\lesssim0.1$ in the figures). The
larger diameter allows us to more accurately determine the weight of the
background curve to be subtracted. At the same time we observe the
distribution of separations is perfectly random outside the region of binary peak.

The probability that the pair in the event is the true binary and not a random
background is given by the ratio%
\begin{equation}
\beta(\hat{\xi})=\frac{P_{bin}(\hat{\xi})}{N_{2}Q(\hat{\xi})}=1-\frac{\gamma
q(\hat{\xi})}{Q(\hat{\xi})}, \label{SA9}%
\end{equation}
where one can replace $\hat{\xi}\rightarrow d_{ij}=2\rho_{2}\hat{\xi}$.
\ Examples of the function $\beta$ will be given below. The shape of the
background distribution $q$ depends on $\hat{\xi}$ only. In the background, we
can see an increase or peak at low $\hat{\xi}$, which is a signal of the
presence of binaries. An additional selection of pairs with small $\alpha
_{ij}$ or a small radial separation $Z_{ij}$ increases the dimension of the
peak on the background, but the shape of the background curve does not change.
For instance, Fig.\ref{ff5}\textit{b,c} corresponds to $\alpha_{ij}%
\leq15\,\mathrm{\deg.}$ The reason is simple: in the sky without binaries
(only background can be observed) there is no correlation between angular
separation $d_{ij}$ and the parameters $\alpha_{ij}$ or $Z_{ij}.$

From the data we have known the numbers of pairs in four windows defined
above: $N_{2}^{A},N_{2}^{B},N_{2}^{C}$ and $N_{2}^{D}$. The number $N_{2}^{A}$
is called the number of binary candidates. According to an assumption above,
we have also numbers of background pairs obtained for $\hat{\xi}\geq2/3$ in
domains%
\begin{equation}
\left\langle 0,\alpha_{c}\right\rangle \rightarrow N_{2}^{I},\qquad
\left\langle \alpha_{c},180\right\rangle \rightarrow N_{2}^{II}. \label{SA10}%
\end{equation}
If we denote%
\begin{equation}
w\left(  x\right)  =\int_{x}^{1}q(\hat{\xi})d\hat{\xi}, \label{SA11}%
\end{equation}
then we can calculate the numbers of background pairs in the windows
\textbf{A-D} as%
\begin{align}
b_{2}^{A}  &  =N_{2}^{I}\frac{1-w\left(  d_{c}\right)  }{w\left(  2/3\right)
},\qquad b_{2}^{B}=N_{2}^{I}\frac{w\left(  d_{c}\right)  }{w\left(
2/3\right)  },\label{SA12}\\
b_{2}^{C}  &  =N_{2}^{II}\frac{w\left(  d_{c}\right)  }{w\left(  2/3\right)
},\qquad b_{2}^{D}=N_{2}^{II}\frac{1-w\left(  d_{c}\right)  }{w\left(
2/3\right)  }.\nonumber
\end{align}
Therefore, the numbers of binaries in the respective windows read%
\begin{align}
n_{2}^{A}  &  =N_{2}^{A}-b_{2}^{A},\qquad n_{2}^{B}=N_{2}^{B}-b_{2}%
^{B},\label{SA13}\\
n_{2}^{C}  &  =N_{2}^{C}-b_{2}^{C},\qquad n_{2}^{D}=N_{2}^{D}-b_{2}%
^{D}.\nonumber
\end{align}
These numbers, obtained in both regions $\mathbf{R}_{1}$ and $\mathbf{R}_{2}$
under different conditions are given in Fig.\ref{ff7}. The numbers in upper panels \textbf{A, B} (the set All events)
follow from panels \textit{b,c} in Fig.\ref{ff5}. The peaks of binaries are
evident, however the area under the background curve is also considerable.
Correspondingly, the quality ratios $\beta^{A}=n_{2}^{A}/N_{2}^{A}$ (in the
second row of the panels \textbf{A)}, representing the probability of the true
binary in window \textbf{A} are not satisfactory, particularly for the region
$\mathbf{R}_{2}$. Parameter $\beta^{A}$\ can be increased by applying the
additional cut on radial separation $\Delta L$. The condition
\begin{equation}
\Delta L\leq\Delta L_{\max}=500\,\mathrm{pc} \label{SA14}%
\end{equation}
applied in the region $\mathbf{R}_{1}$ gives \textquotedblleft
enriched\textquotedblright\ sample with the quality ratio $\beta^{A}=0.77$
(left middle panel). In the region $\mathbf{R}_{2}$ the situation is more
complicated. The density of stars is higher, so the background grows.
Moreover, the average density varies with longitude. So, we have divided the
region into three subregions with different $\Delta L_{\max}$ cuts, which give
a similar $\beta^{A}$ in these subregions. Their definition is shown in
Tab.\ref{Tab3}. \begin{table}[ptb]
\begin{center}%
\begin{tabular}
[c]{cccc}%
$l$[deg] & $\left\langle -30,+30\right\rangle $ & $\left\langle \pm
30,\pm90\right\rangle $ & $\left\langle 90,270\right\rangle $\\\hline
$\Delta L_{\max}[\mathrm{pc}]$ & 50 & 100 & 400
\end{tabular}
\end{center}
\caption{Cuts on radial separation in galactic longitude subregions of region
\textbf{R}$_{2}.$}%
\label{Tab3}%
\end{table}These cuts give resulting $\beta^{A}$ and other parameters related
to $\mathbf{R}_{2}$ listed in right middle panel. One could further squeeze
the cuts to obtain a cleaner sample of binaries (higher ratio $n_{2}^{A}%
/N_{2}^{A}$), however at the price that more true binaries are excluded
($n_{2}^{A}\ $\ is smaller). This is illustrated by the numbers in upper and
middle yellow panels in the figure. The cuts exclude not only background pairs
but also true binaries in window \textbf{A}. It is due to a rather poor
accuracy of radial separation that we have shown in Fig.\ref{ff4}\textit{b}.
If we had an accurate radial separation, then a suitable cut would suppress
only background and preserve true binaries.

\begin{figure}[t]
\centering\includegraphics[width=80mm]{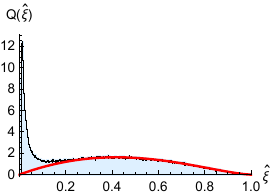}\caption{Scaled separation
$\hat{\xi}$: peak of binaries of magnitude $G$ $\leq15$ and random background
(\textit{red curve}) in region $\mathbf{R}$.}%
\label{ff8}%
\end{figure}

The situation is more favorable with brighter stars. The lower panels in
Fig.\ref{ff7} show that for binaries of magnitude $G$ $\leq15$ the quality
ratio $\beta^{A}$ is very good even without any cut on radial separation. The
quality of this sample is illustrated also by Fig.\ref{ff8}, where we observe
the high binary peaks with low background. In any case, our criterion is based on the resulting ratio
$\beta^{A}$, so we can accept candidates with greater parallax uncertainty if
$\beta^{A}$ is greater than requested. In other words, if the signature from
$d_{ij}$ and $\alpha_{ij}$ is sufficient, then the parallax is not important.

\begin{figure}[t]
\centering\includegraphics[width=75mm]{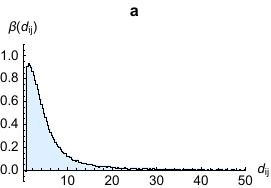}\includegraphics[width=75mm]{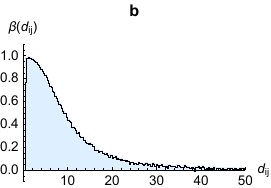}
\includegraphics[width=75mm]{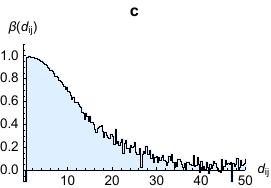}\includegraphics[width=75mm]{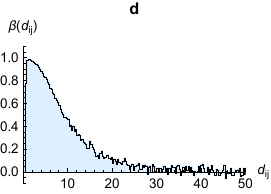}\caption{Function
$\beta$ defines probability of true binary at different separations in
$\mathbf{R}$ for $\alpha\leq\alpha_{c}.$ Panel\textit{\ a:} all events.
Panel\textit{\ b:} enriched exposition.\ Panel\textit{\ c:} only stars of
magnitude $G\leq15.$ Panel \textit{d:} enriched exposition, event subsample
$\rho_{2}=36$as. \ Unit: $d_{ij}[\mathrm{as}].$}%
\label{ff6}%
\end{figure}

Equivalently, abundance of binaries depending on their separation is defined
by $\beta$ function (\ref{SA9}). In Fig.\ref{ff6}\textit{a,b,c} we have shown
this function for some subsets of regions $\mathbf{R}_{1}$ and $\mathbf{R}%
_{2}$ defined by Fig.\ref{ff7}. Figure Fig.\ref{ff6}\ demonstrates that higher $\beta^{A}$ could
be also reached by squeezing the parameter $d_{c}$ in window \textbf{A}
(\ref{SA4}). Note, the effectively wider peak for magnitude $G$ $\leq15$ is
due to a lower level of background pairs, which can be seen from the
comparison of Fig.\ref{ff8} with Fig.\ref{ff5}\textit{b,c}.

We have shown that the function $\beta$ is a key to determining the quality
and quantity of binaries in the selected subset. It is, therefore, necessary
to verify that this function does not change with the choice of the purely
technical parameter $\rho_{2}$. Fig.\ref{ff6}\textit{d} shows the function
$\beta$ calculated for the sub-sample of enriched exposition with event radius
$\rho_{2}=36$as (half the radius of standard events). The functions in panels
\textit{b,d} agree perfectly. A similar agreement was also verified for
Figs.\ref{ff9} - \ref{ff11} below.

\begin{figure}[t]
\centering\includegraphics[width=80mm]{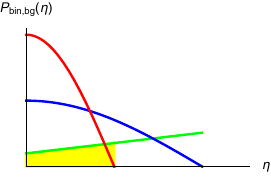}\caption{Distributions of
true binaries $P_{bin}$ (\textit{red }and\textit{ blue, }which is wider, due
to measurement errors) and background $P_{bg}$ (\textit{green}). Parameter
$\eta$ represent some selection parameter $(d_{ij},\alpha_{ij},...)$. The
numbers of the true binaries $n_{2}$ are given by the area under \textit{red}
and \textit{blue} curves. Corresponding numbers of background pairs $b_{2}$
are given by the \textit{yellow} area and the whole area under the
\textit{green} curve. }%
\label{ff20}%
\end{figure}

What is the effect of input data uncertainties on the selection quality? The
initial distribution $P(d_{ij},\alpha_{ij})$ \ in Fig.\ref{ff3} consists of
two parts
\begin{equation}
P(d_{ij},\alpha_{ij})=P_{bin}(d_{ij},\alpha_{ij})+P_{bg}(d_{ij},\alpha_{ij}),
\label{SA31}%
\end{equation}
corresponding to the true binaries and the background. The shape of the
distribution $P_{bin}(d_{ij},\alpha_{ij})$ is affected by the measurement
errors. In general, these errors cause a widening of this distribution, but
its integral $n_{2}$ does not change, as suggested in Fig.\ref{ff20}.
However, the errors will affect the value of the quality ratio%
\begin{equation}
\beta^{P}=\frac{n_{2}}{n_{2}+b_{2}} \label{SA32}%
\end{equation}
in the domain of the peak. As suggested in the figure, wider $P_{bin}$ means
larger $b_{2},$ which implies a smaller ratio $\beta^{P}$. In this way, the
measurement errors of the selection parameters directly affect the $\beta$
function, larger errors mean worse $\beta$. At the same time, larger errors
expand the space for selection. A similar result follows from the algorithm in
(\cite{halb}) for the selection of common proper motion pairs.

At the end of this section, we summarize the selection algorithm of binary candidates:

1. In the distribution $P(d_{ij},\alpha_{ij})$ \ in Fig.\ref{ff3} we observe
the peak on a smooth background. The peak is a sign of the presence of
binaries. We take the edges (\ref{SA0a}) of this peak as the limits of our
selection of\ binary candidates.

2. The number of true binaries can be accurately determined by subtracting the
random background as shown in Fig.\ref{ff5}\textit{b,c}. The background curve
is given by the function (\ref{SA23}). This curve corresponds to the random
distribution of sources within a circle. That is why we work with
\textit{circular events} and with the distribution of \textit{angular
separations} of the pairs.

3. The background level in the peak domain can be further reduced by the cuts
on radial separation $\Delta L_{\max}$.

4. The cuts $\alpha_{c}$ and $\Delta L_{\max}$ reduce the background level,
but does not change the shape of the corresponding background curve
$q(\hat{\xi})$. This distribution, after the renormalization (\ref{SA0z}), is
used for calculation of the probabilistic function $\beta$, which defines the
selection quality.

\subsection{Projected absolute separations, periods and masses of binaries}

\label{pasp}
\begin{figure}[t]
\centering\includegraphics[width=80mm]{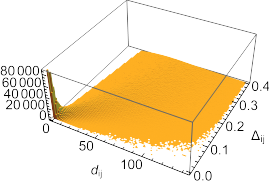}\includegraphics[width=80mm]{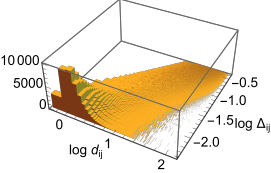}\caption{Correlation
of angular separation $d_{ij}[\mathrm{as}]$\ and \ projected distance
$\Delta_{ij}[\mathrm{pc}]$ for $\alpha\leq\alpha_{c}.$ Left a right panel -
linear and logarithmic scales. Binning: $1.44$as$\times0.004$pc,
$0.36$as$\times0.001$pc.}%
\label{ff9}%
\end{figure}

\begin{figure}[t]
\centering\includegraphics[width=55mm]{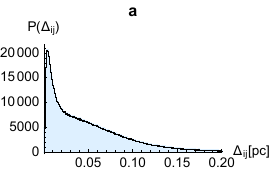}\includegraphics[width=55mm]{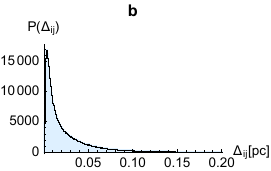}
\includegraphics[width=55mm]{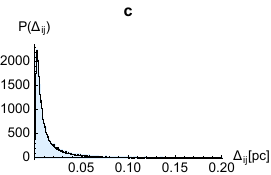}\caption{Projected distance
$\Delta_{ij}$ of pairs in domains \textbf{A} of regions $\mathbf{R}_{1}$ and
$\mathbf{R}_{2}$\textbf{\ }defined in Fig.\ref{ff7}. Panel\textit{\ a:} all
events, Panel\textit{\ b:} enriched exposition, Panel\textit{\ c:} only stars
of magnitude $G\leq15$. Binning: 0.0004pc.}%
\label{ff10}%
\end{figure}

\begin{figure}[t]
\centering\includegraphics[width=55mm]{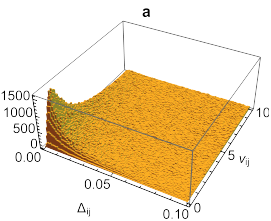}\includegraphics[width=55mm]{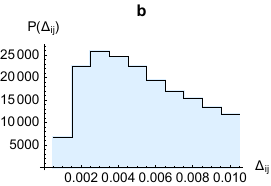}
\includegraphics[width=55mm]{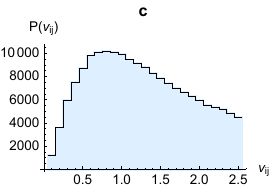}\caption{Panel\textit{\ a:
}Correlation of the transverse separation $\Delta_{ij}$ with the transverse
velocity $v_{ij}$ of orbital motion in region $\mathbf{R}$.
Panels\textit{\ b,c: }Distributions of $\Delta_{ij}$ and $v_{ij}$ in domain
(\ref{SA2}). Units: $\Delta_{ij}[\mathrm{pc}],v_{ij}[\mathrm{km/s}].$ Binning:
$0.001$pc$\times0.1$km/s$,0.001$pc$,0.1$km/s.}%
\label{ff11}%
\end{figure}

In Fig.\ref{ff9} we have shown an important correlation in the peak of binaries,
where angular separation $d_{ij},$ as one would expect, strongly correlate
with the projected distance $\Delta_{ij}$ that is \ calculated as%
\begin{equation}
\Delta_{ij}=d_{ij}\frac{L_{i}+L_{j}}{2}. \label{SA14a}%
\end{equation}
Distribution of $\Delta_{ij}$ in the region of peak (windows \textbf{A} in
Fig.\ref{ff7}) is shown in Fig.\ref{ff10}. In panels \textit{b,c} we have shown distributions from subsets
with higher rate of binaries. We observe that%
\begin{equation}
\Delta_{ij}\lesssim0.1\,\mathrm{pc}, \label{SA15}%
\end{equation}
which confirms our former result from 3D analysis. In panel \textit{a} we
observe presence of pairs, which contradicts this constraint. It is due to
high rate of background pairs in this set. Note that nearby binaries with
wider separation can be outside of our acceptance window \textbf{A}. In fact
only candidates with separation%
\begin{equation}
\Delta_{ij}\leq d_{c}L\approx7\times10^{-5}L, \label{SA19}%
\end{equation}
where $L$ is average distance of the pair, are accepted.

\ \ Similarly as before (\cite{AJpzkp}), we can estimate the projection of the
orbital velocity of the binary as
\begin{equation}
v_{ij}=\left\vert \mathbf{u}_{i}-\mathbf{u}_{j}\right\vert \frac{L_{i}+L_{j}%
}{2}. \label{SA1}%
\end{equation}
From the plot in Fig.\ref{ff11}\textit{a}, we can roughly estimate averages of the orbital periods and total
masses of the binary systems. The binaries are accumulated in a peak very
similar to that obtained earlier in 3D analysis. Very similar plot (with much
lower statistics) can be obtained also for the subset of binaries with
magnitude $G$ $\leq15$. If we take the sources roughly in the domain of
half-width of the maximum (panels\textit{\ b,c})
\begin{equation}
\Delta_{ij}\leq0.01\,\mathrm{pc},\qquad\nu_{ij}\leq2.5\mathrm{km/s}
\label{SA2}%
\end{equation}
then in the approach described in (\cite{AJpzkp}) we now obtain%
\begin{equation}
\left\langle T\right\rangle \approx4.2\times10^{4}\mathrm{y,\qquad
}\left\langle M_{tot}\right\rangle \approx0.65\mathrm{M}_{\odot}, \label{SA3}%
\end{equation}
which is comparable with the previous approximate estimate ($8\times
10^{4}\mathrm{y,}$ $0.8\mathrm{M}_{\odot}$) obtained in the cited paper. These
numbers can also be compared with the results of the sample of thirty CPM
(Common Proper Motion) pairs, which were discussed in (\cite{duqu}), Table 4.
These pairs represent wide binaries in the solar neighborhood, for which the
table lists the parameters $\log T_{i}$ and masses $M_{1i}$ and $M_{2i}$. We
have calculated the mean values%
\begin{equation}
\left\langle T\right\rangle =10^{\left\langle \log T_{i}\right\rangle }%
\approx2.6\times10^{4}\mathrm{y,\qquad}\left\langle M_{tot}\right\rangle
=\left\langle M_{1i}+M_{2i}\right\rangle \approx1.6\mathrm{M}_{\odot},
\label{SA33}%
\end{equation}
which are also well comparable with our estimates above.

\subsection{Wide trinaries and quaternaries}

\label{witq}So far we have assumed the excess of close pairs (window
\textbf{A}) is due to only \ binaries, which is correct only in a first
approximation. A more detailed analysis indicates also a limited presence of
multiple systems, trinaries and quaternaries. To identify their candidates, we
will generalize our criterion of binary candidates. For each pair in the
candidate multiple system we require the same condition as for separation of
binaries (\ref{SA4}). Three stars create three separations, four stars create
six separations. In Tab.\ref{Tab4} \begin{table}[ptb]
\begin{center}%
\begin{tabular}
[c]{cccccccccc}
&  & \multicolumn{4}{c}{enriched exposition} & \multicolumn{4}{c}{$G$ $\leq
15$}\\
region & $m$ & $N_{m}^{A}$ & $b_{m}^{A}$ & $n_{m}^{A}$ & $\beta$ & $N_{m}^{A}$
& $b_{m}^{A}$ & $n_{m}^{A}$ & $\beta$\\\hline\hline
R$_{1}$ & $3$ & $1,724$ & $137$ & $1,587$ & $0.921$ & $53$ & $2$ & $51$ &
$0.96$\\
& $4$ & $12$ & $0$ & $12$ & $1.$ & $1$ & $0$ & $1$ & $1.$\\\hline
R$_{2}$ & $3$ & $1,946$ & $147$ & $1,789$ & $0.924$ & $119$ & $5$ & $114$ &
$0.96$\\
& $4$ & $10$ & $0$ & $10$ & $1.$ & $0$ & $0$ & $0$ & $x$\\\hline
$\Sigma$ & $3$ & $3,670$ & $284$ & $3,386$ & $0.923$ & $172$ & $7$ & $165$ &
$0.96$\\
& $4$ & $22$ & $0$ & $22$ & $1.$ & $1$ & $0$ & $1$ & $1.$%
\end{tabular}
\end{center}
\caption{Statistics of trinary and quaternary candidates ($m=3,4$) in regions
$\mathbf{R}_{1}$ and $\mathbf{R}_{2}$. Table shows the numbers of candidates,
estimated background and true trinaries and quaternaries ($N_{m}^{A}$,
$b_{m}^{A}$, $n_{m}^{A}$), $\beta=n_{m}^{A}/N_{m}^{A}.$}%
\label{Tab4}%
\end{table}we have shown statistics of events involving just one isolated
trinary or quaternary candidate $(N_{m}^{A})$. The random background events
$(b_{m}^{A})$ are obtained from the real events in which the local coordinates
$\{x_{i},y_{i}\}$ are replaced by random positions inside the same events. The
difference $N_{m}^{A}-b_{m}^{A}$ represents the estimate of number of the true
multiple systems $n_{m}^{A}$. The presence of multiple systems implies that
the number of binary candidates $N_{2}^{A}$ should be reduced by%
\begin{equation}
\Delta N_{2}^{A}=3N_{3}^{A}+6N_{4}^{A}, \label{SA16}%
\end{equation}
which gives $\approx1.6\%$ in the set of enriched exposition.

\begin{figure}[t]
\centering\includegraphics[width=75mm]{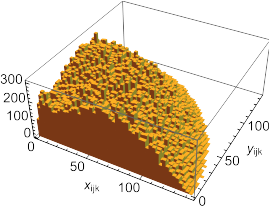}\includegraphics[width=75mm]{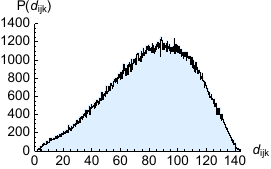}
\includegraphics[width=75mm]{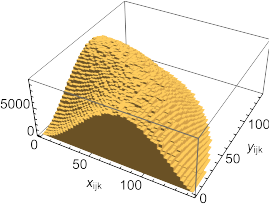}
\includegraphics[width=75mm]{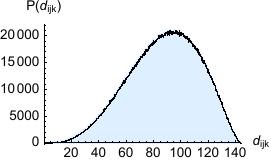}\caption{Distributions of angular
separations $x_{ijk},y_{ijk}$ and $d_{ijk}$ in trinary candidates. Upper
panels show results from the region $\mathbf{R}$ (enriched exposition). The
lower panels represent background. Unit: $x_{ijk},y_{ijk},d_{ijk}%
[\mathrm{as}].$ Binning: $2.88$as$\times2.88$as$,0.288$as.}%
\label{ff12}%
\end{figure}

\begin{figure}[t]
\centering
\includegraphics[width=75mm]{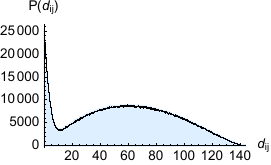}\includegraphics[width=75mm]{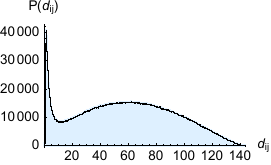}
\caption{Distributions of angular separations $d_{ij}$. \textit{Left:} Random
sources with the admixture of close pairs. \textit{Right:} Real event data
from region $\mathbf{R}$ (enriched exposition). Unit: $d_{ij}[\mathrm{as}].$
Binning: $0.288$as.}%
\label{ff13}%
\end{figure}

\begin{figure}[t]
\centering\includegraphics[width=75mm]{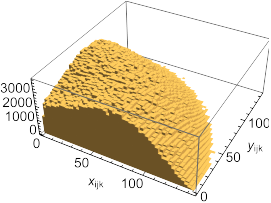}\includegraphics[width=75mm]{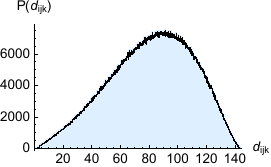}\caption{Distributions
of angular triangle separations $x_{ijk},y_{ijk}$ and $d_{ijk}$ for the same
set of simulated events used in the previous figure. Unit: $x_{ijk}%
,y_{ijk},d_{ijk}[\mathrm{as}].$ Binning: $2.88$as$\times2.88$as$,0.288$as.}%
\label{ff14}%
\end{figure}

For each triplet of stars inside the event we define the triangle separation
as%
\begin{equation}
d_{ijk}=\max(d_{ij},d_{jk},d_{ki}). \label{SA17}%
\end{equation}
If $\ \alpha\beta$ are subscripts of the most separated pair, then the
components of triangle separation are defined as%
\begin{equation}
x_{ijk}=\left\vert x_{\alpha}-x_{\beta}\right\vert ,\qquad y_{ijk}=\left\vert
y_{\alpha}-y_{\beta}\right\vert . \label{SA20}%
\end{equation}
Distributions of these parameters are shown in Fig.\ref{ff12}
together with the distributions corresponding to the background,
which is generated by randomly distributed sources in the event. The high
level of the background under the slight trinary peak at $d_{ijk}$
$\lesssim20\,\mathrm{as}$ (upper panel left) is due to primarily by binaries.
Each binary generates also excess of triangle separations. It is illustrated
in Figs.\ref{ff13} and \ref{ff14}. In the first figure (left panel) we show the
simulated distribution of separations of random sources with the admixture of
extra close pairs. The parameters of this admixture (\textit{probability} of
close pairs and \textit{width} of normal distribution of their separations)
are set to reproduce distributions in the real events (right panel).
Distributions of triangle separations obtained from the same set are shown in
Fig.\ref{ff14}. The last figure explains a high background under the trinary peak
in upper panels of Fig.\ref{ff12}.

Fig.\ref{ff7} and Tab.\ref{Tab4} show that in the window \textbf{A}%
($\mathbf{R}$) of enriched exposition we have $511,998$ binaries, $3,404$
trinaries and $22$ quaternaries. However, the total numbers are greater. The
estimate based on the all events window (\textbf{A-D}) in upper panels of
\ Fig.\ref{ff7} gives the total number of binaries roughly $1.5\times10^{6}$.
However, as we discussed in (\cite{AJpzkp}), part of them (mainly in windows
\textbf{B,C}) may be an image of widening pairs that were less separated but
weakly bound in the past.

\section{Catalogue}

\label{SEC3}In this section, we describe the catalogue of the multiple star
candidates, which are selected with the use of the events defined by
Tab.\ref{Tab1}. For the present version of the catalogue, we accept only the
candidates from window \textbf{A}($\mathbf{R}$) of the enriched exposition
defined in the middle part of Fig.\ref{ff7}. So, the selected pairs satisfy
the conditions of angular separation, collinearity and radial separation:
\begin{equation}
d\leq15\,\mathrm{as},\qquad\alpha\leq15\deg,\qquad\Delta L\leq\Delta L_{\max},
\label{sabe37}%
\end{equation}
where $\Delta L_{\max}$ is defined in relation (\ref{SA14}) for region
$\mathbf{R}_{1}$ and in Tab.\ref{Tab3} for $\mathbf{R}_{2}$. We record:

1. Binary candidates: separate pairs that satisfies (\ref{sabe37}).

2. Trinary candidates: separate triplets of sources, where each pair (3 in
total) satisfies (\ref{sabe37}).

3. Quaternary candidates: separate quaternions of sources, where each pair (6
in total) satisfies (\ref{sabe37}).

\begin{figure}[t]
\centering\includegraphics[width=27mm]{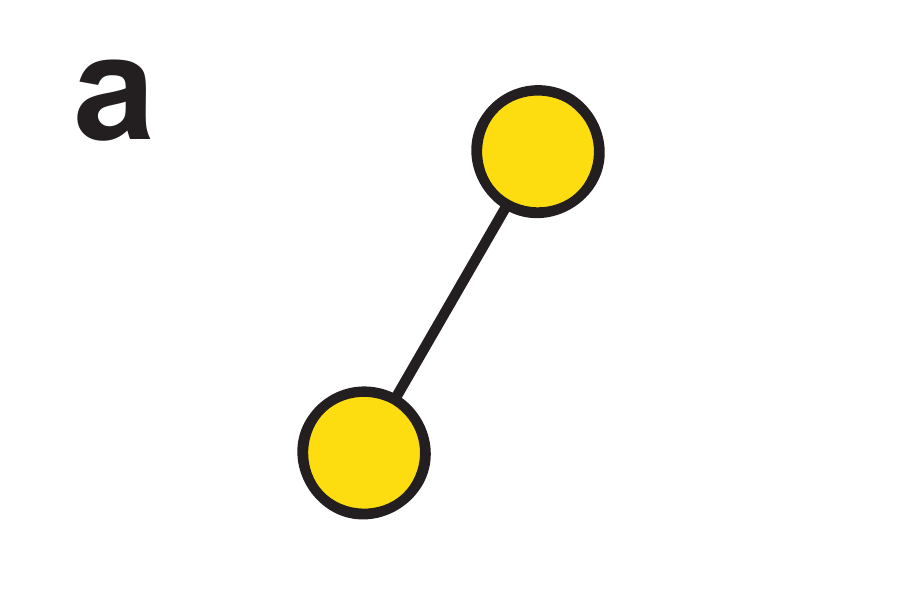}\includegraphics[width=27mm]{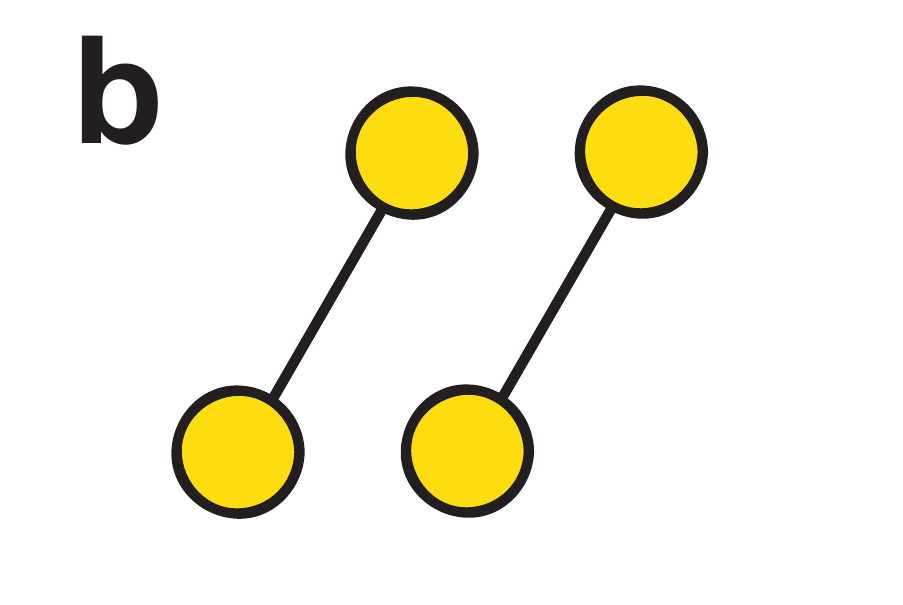}
\includegraphics[width=27mm]{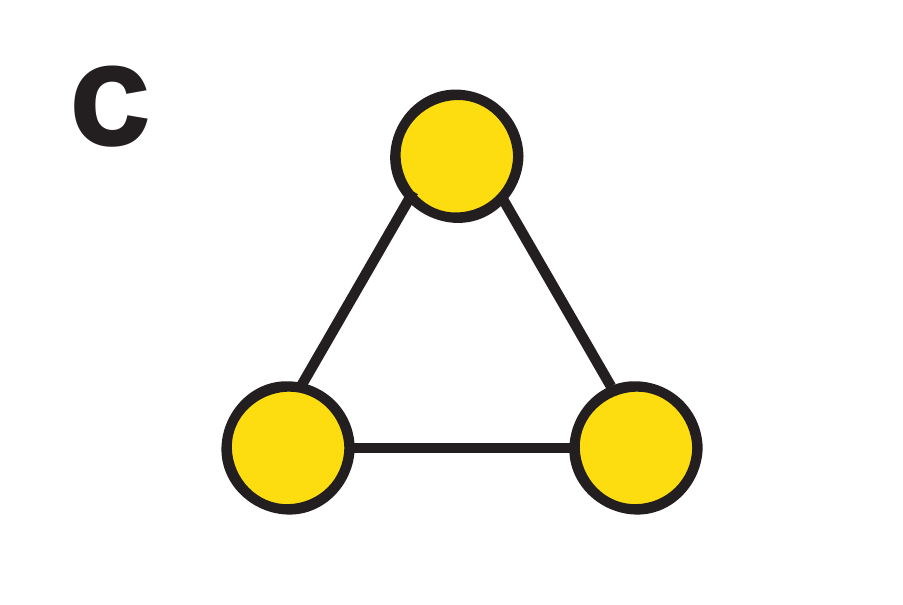}\includegraphics[width=27mm]{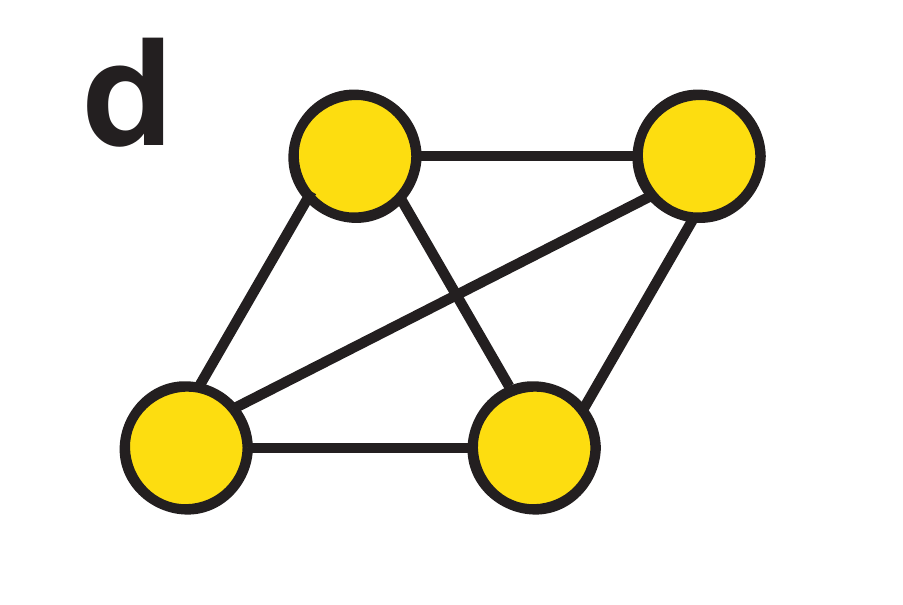}\includegraphics[width=27mm]{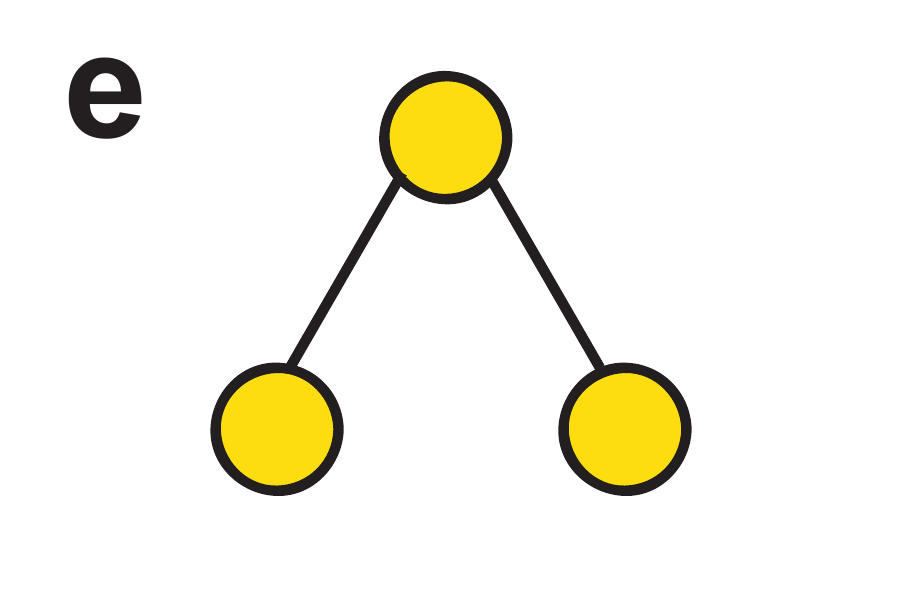}
\includegraphics[width=27mm]{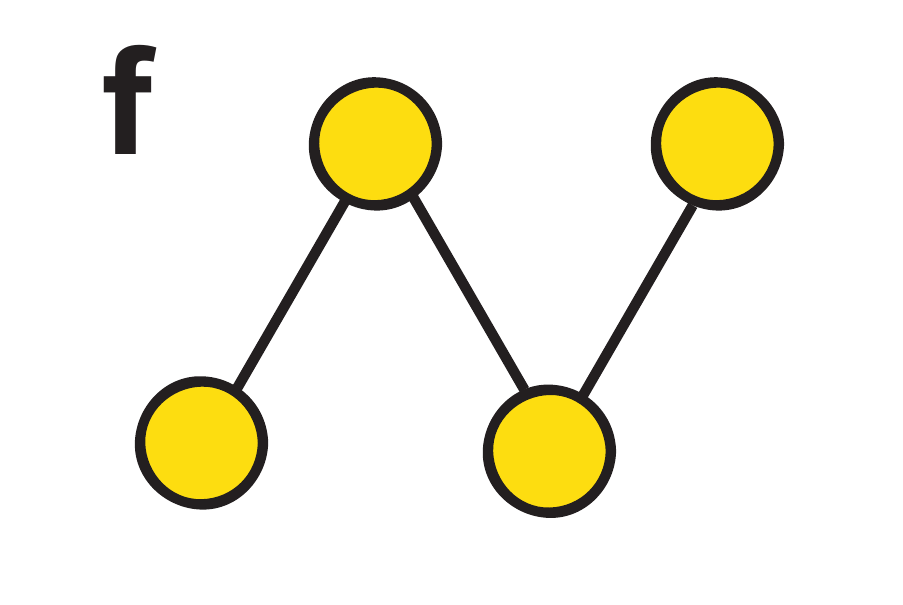}\caption{Examples of patterns relevant
to the selection (rejection) of candidates for multiple stars.}%
\label{ff17}%
\end{figure}

Examples of possible patterns in window \textbf{A }generated in one event are
symbolically shown in Fig.\ref{ff17}. We skip the events with empty window \textbf{A.} Pairs with the
bar meet (\ref{sabe37}), pairs without the bar do not. Now we work with the
patterns, where each source has at least one bar. According to the above
definition, diagrams represent the candidates in the event: one binary
(\textit{a}), two binaries (\textit{b}), one trinary (\textit{c}), one
quaternary (\textit{d}). The rate of other combinations of candidates is
negligible. Combinations that do not correspond to the defined candidates,
such as diagrams \textit{e, f}, also have very little weight. We skip them and
accept only candidates \textit{a-d} into the catalogue. The circle events
cover only part of the sky (corresponding to the fraction $\pi/4$). We also
lose candidates between neighboring events when the pairs are split between
the neighbors. To recover these losses, we work with modified coverage:

\textit{i)} The event circles of radius $72\,\mathrm{as}$ are replaced by
squares of edge $144\mathrm{as}$ with no gaps between them. In each square, we
search for multiple star candidates.

\begin{figure}[t]
\centering\includegraphics[width=80mm]{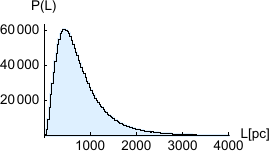} \caption{Distances of all
candidate sources. Binning: $30$pc.}%
\label{ff15}%
\end{figure}

\textit{ii)} The procedure is repeated with the same squares centred in the
corners and the edge centres of the former squares (we have $4$ grids in
total). Then the search results are merged, the summary data are given in
Tab.\ref{tab5}. \begin{table}[ptb]
\begin{center}%
\begin{tabular}
[c]{ccccc}%
$m$ & $N_{m}^{A}$ & $\beta$ & $n_{m}^{A}$ & $\Delta n_{m}^{A}$\\\hline
$2$ & $900,842$ & $0.733$ & $660,317$ & $696$\\
$3$ & $5,282$ & $0.923$ & $4,875$ & $67$\\
$4$ & $30$ & $1.$ & $30$ & $6$%
\end{tabular}
\end{center}
\caption{Summary table of multiple stars in the window \textbf{A}($\mathbf{R}%
$) of enriched exposition. }%
\label{tab5}%
\end{table}The quality ratios $\beta=n_{m}^{A}/N_{m}^{A}$ are taken from the
analysis of circle events with well defined background (Fig.\ref{ff7} and
Tab.\ref{Tab4}). Now we can use them to estimate $n_{m}^{A}$ and corresponding
statistical errors%
\begin{equation}
\Delta n_{m}^{A}=\beta\sqrt{N_{m}^{A}}. \label{SA25}%
\end{equation}
It is seen the number of multiple systems decreases rapidly with their
multiplicity:%
\begin{equation}
\frac{n_{3}^{A}}{n_{2}^{A}}\approx0.7\%,\qquad\frac{n_{4}^{A}}{n_{3}^{A}%
}\approx0.6\%. \label{SA18}%
\end{equation}
In Fig.\ref{ff15} we have shown distribution of distances of all stars accepted to
the catalog. Note that selection criteria exclude preferably the most distant
sources from distribution in Fig.\ref{ff1}. Obviously, the candidates of
higher quality (but lower quantity) can be obtained from the catalog by
reselection with more strict cuts (\ref{sabe37}).

The catalogue of selected candidates is represented by a matrix, which is
defined as follows. Each row represents one star and there are the following
data in the columns:

1--2: \ Group ID and Group size $\left(  n=2,3,4\right)  $ to match stars with
the group they belong to.

3--96: \ Copy of the original entry for the star from \textit{Gaia}-DR2
catalogue
\footnote{\url{http://cdn.gea.esac.esa.int/Gaia/gdr2/gaia_source/csv/}},
according to the documentation
\footnote{\url{http://gea.esac.esa.int/archive/documentation/GDR2/Gaia_archive/chap_datamodel/sec_dm_main_tables/ssec_dm_gaia_source.html}}%
.

97--98: \ Minimum and maximum angular separation of the star from other stars
in the group [$\mathrm{as}$].

99--100: \ Minimum and maximum projected physical separation (\ref{SA14a}) of
the star from other stars in the group [pc].

\begin{figure}[t]
\centering\includegraphics[width=80mm]{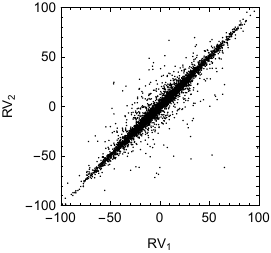} \caption{Correlation of
radial velocities for 6469 pairs. Unit: [km/s].}%
\label{ff18}%
\end{figure}

The quality of our catalogue can be easily verified for brighter binaries,
where the DR2 data includes radial velocities. The catalogue contains 6469
such pairs, only 16 of them involves a star with magnitude $G>15$. In
Fig.\ref{ff18} we show the correlation of both velocities, which confirms the
common radial motion of the pairs. This result agrees very well with the same
correlation obtained from the DR2 catalogue in (\cite{andrews}) and
(\cite{sapo}).

\section{Comparison with other catalogues}

\label{SEC4}The catalogues of wide binaries (\cite{badry,hart,esteban,AJpzkp})
and our present one are based on the selection defined by three parameters
measured by the \textit{Gaia:} angular position (separation), proper motion
and parallax. However, catalogues differ in the choice of algorithm, which
processes these variables. To illustrate, we will compare the algorithm used
in this study (A1) with the algorithm used for the recent catalogue
(\cite{badry}) containing a comparable amount of binaries (A2).

\textit{i) angular separation}

This parameter is measured with a high accuracy. The presence of binaries
generates a clear peak in its distribution at small separations. The random
background in the events is exactly described by the function (\ref{SA23}).
The peak in angular separations is the basic signature of binaries in the
algorithm A1.

From angular separation one can, with the use of parallax, calculate projected
separation. Both parameters are strongly correlated. The cut on projected
separations $\Delta\leq1$pc is applied in the algorithm A2. However, the
question of the background is here more complex, see discussion below.

\textit{ii) proper motion}

\begin{figure}[t]
\centering\includegraphics[width=80mm]{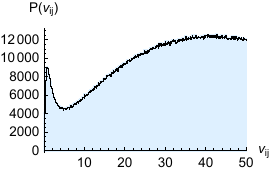}\includegraphics[width=80mm]{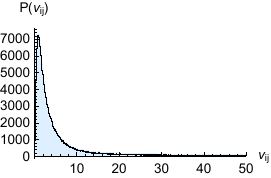}
\caption{Distribution of velocities $v_{ij}$ for all pairs \textit{(left)} and
for pairs in the peak (\ref{SA0a}) \textit{(right)} in the region
\textbf{R}$_{1}$ (enriched exposition). Unit: $v_{ij}$[km/s]. Binning: 0.1
km/s.}%
\label{ff16}%
\end{figure}

In algorithm A2, the difference of velocities $v_{ij}$ (\ref{SA1}) is limited
by the cut
\begin{equation}
v_{ij}\leq v_{\max}. \label{SA26}%
\end{equation}
The algorithm A1 is based on conditions (\ref{SA0a}), which define the domain
of the peak with the high population of binaries (Fig.\ref{ff3}). The
algorithm includes an accurate calculation of the background under the peak.
What are the velocities $v_{ij}$ inside and outside this peak? The answer is
given in Fig.\ref{ff16}. The left panel shows the distribution of velocities for all pairs
within the event, the peak at small $v_{ij}$ corresponds to the binaries. The
right panel shows the same distribution for the pairs in the peak domain.
Corresponding pairs are mostly binaries, so the large velocities are strongly
suppressed. The panel represents the distribution of the orbital motion
projection. One can admit, the tail of distribution may still be contaminated
with some background pairs. Alternatively, instead of our cut
\begin{equation}
\alpha_{ij}\leq\alpha_{\max} \label{SA27}%
\end{equation}
it would be possible to set the cut (\ref{SA26}). In the present study, we
preferred (\ref{SA27}), because the angle $\alpha_{ij}$ does not depend on the
parallax, which has a large measurement error. For future analysis of the
\textit{Gaia} DR3 data, in which a more accurate parallax measurement is
expected, we plan to try replacing (\ref{SA27}) with (\ref{SA26}), which can
more effectively suppress the background.

\textit{iii) parallax}

The parallax $p$ or the distance $L=1/p$ are determined with a large error,
$\Delta L$ is usually larger than the expected binary dimension. Large radial
separation does not necessarily mean that the pair is not true binary.
However, a smaller radial separation (or difference of parallaxes) increases
the probability that the pair is true binary. Therefore the algorithm A1
requires limited radial separation defined by (\ref{SA23}) and Tab.\ref{Tab3}.
Similarly, the A2 requires limited difference of parallaxes.

Finally, both algorithms similarly reject denser clusters of sources: event
multiplicity $M>25$ (A1) and number of neighbors $>30$ (A2). In general,
algorithms have a similar philosophy but differ in technical details.

A very important technical step is to define random background (A1) or
equivalently the chance alignments (A2). In this respect, the two algorithms
differ significantly. Let's make a comparison.

A1: The distribution of separations of random sources inside a circle is given
exactly by formula (\ref{SA23}). The shape of this curve does not depend on
other parameters such as magnitude, the direction of proper motion or the
parallax. The binary peak can be separated from the background described by
this curve, see Sec.\ref{sebi} and examples in Figs.\ref{ff6} and \ref{ff6}.
The normalized background curve is
\begin{equation}
q\left(  \hat{\xi}\right)  \allowbreak\approx8\hat{\xi}+... \label{SA29}%
\end{equation}
for small $\hat{\xi}$ (peak region). The probability of finding binary in a
selected subset is given by function $\beta$ defined by ratio (\ref{SA9}),
examples are shown in Fig.\ref{ff6}.

A2: From the \textit{Gaia} EDR3 data input, there are produced two files
involving pairs with seven parameters $\mathbf{x}=\{$angular separation,
distance, parallax difference uncertainty,...$\}$:

a) Catalogue of candidates

b) Catalogue of chance alignments (shifted catalogue)

The densities of pairs $N_{\text{candidates}}(\mathbf{x})$ and
$N_{\text{chance align}}(\mathbf{x})$ in the seven-dimensional parameter space
are approximated by the Gaussian kernel density estimates. The ratio of these approximations%

\begin{equation}
\mathcal{R}(\mathbf{x})=\frac{N_{\text{chance align}}(\mathbf{x}%
)}{N_{\text{candidates}}(\mathbf{x})} \label{SA30}%
\end{equation}
is the parameter, which provides classification of the quality of candidates:
the low $\mathcal{R}$ means a high probability that the candidate is true
binary, as illustrated in Fig. 5 in \cite{badry}. In the right panel of this
figure, we observe: the lower $\mathcal{R}$ (and higher quality of the
candidate) means stronger suppression of more separated pairs. It is the
result, which correlates with the shape of probabilistic function $\beta$ in
A1. However, the ratio $\mathcal{R}$ of both approximations does not strictly
mean \textit{probability}, which the authors admit.

We will compare the contents of the catalogues listed in Tab.\ref{tab6}.
\begin{table}[ptb]
\begin{center}%
\begin{tabular}
[c]{l|ccc}
& $N_{tot}$\  & $N_{b>25}$ & reference/DR\\\hline
A1 & \multicolumn{1}{|l}{$900,842$} & $900,842$ & this paper/DR2\\
A2 & \multicolumn{1}{|l}{$1,256,400$} & $496,888$ & \cite{badry}/EDR3\\
A3 & \multicolumn{1}{|l}{$93,898$} & $55,319$ & \cite{hart}/DR2\\
A4 & \multicolumn{1}{|l}{$80,560$} & $40,107$ & \cite{AJpzkp}/DR2\\
A5 & \multicolumn{1}{|l}{$3,055$} & $381$ & \cite{esteban}/DR2\\
A6 & $9,977$ & $5,546$ & \cite{sapo}/DR2
\end{tabular}
\end{center}
\caption{Numbers of the binary candidates in the compared catalogues.}%
\label{tab6}%
\end{table}We define three types of candidates for any two compared catalogues
A$i$, A$j$:

1) unique - the candidate appears in only one catalogue $\left(  N_{i}%
^{uniq}\right)  $

2) identical - the candidate appears in both catalogues $\left(  N_{ij}%
^{iden}\right)  $

3) indefinite - cannot be decided, for example, two candidates from two
catalogues have only one common star, or the pair in one catalogue is part of
a greater system in another one.

The results of the comparison are shown in Tab.\ref{tab7} \begin{table}[ptb]
\begin{center}%
\begin{tabular}
[c]{c|cccccc}
& A1 & A2 & A3 & A4 & A5 & A6\\\hline
A1 &  & $343,302$ & $39,031$ & $24,949$ & $59$ & $2,862$\\
A2 & $151,738\backslash555,691$ &  & x & x & x & x\\
A3 & $54,530\backslash861,469$ & x &  & $28,312$ & $85$ & $7,290$\\
A4 & $54,707\backslash874,969$ & x & $51,811\backslash65,148$ &  & $108$ &
$7,937$\\
A5 & $2,979\backslash900,765$ & x & $2,967\backslash93,810$ & $2,937\backslash
80,442$ &  & $11$\\
A6 & $7,115\backslash897,980$ & x & $7115\backslash897980$ & $2,040\backslash
72,693$ & $9966\backslash3015$ &
\end{tabular}
\end{center}
\caption{Comparison of catalogues A1 -A6 shows the numbers of identical
$N_{ij}^{iden}$ (above diagonal) and unique $N_{i}^{uniq}\backslash
N_{j}^{uniq}$ (below diagonal) candidates.}%
\label{tab7}%
\end{table}In the section above the diagonal, there are the numbers
$N_{ij}^{iden}$, below the diagonal we have the numbers $N_{i}^{uniq}%
\backslash N_{j}^{uniq}$. In the catalogue A2, only such pairs can be used for
comparison, where the IDs of both stars appear also in the\textit{ Gaia} DR2
data. We perform the comparison of A2 only with our catalogue (A1, $\left\vert
b\right\vert >25\deg$), so the other corresponding places in the table are
empty (x). Unique candidates are based on different selection conditions in A1
and A2. For example, in A2, the unique part is generated mostly by candidates
with greater angular separation ($d>15\,\mathrm{as}$) than is accepted in A1.
On the other hand, A1 imposes much weaker constraints on parallaxes, which
generate its unique part. That is why in Fig.\ref{ff15} we observe many
candidates more distant than $1000$pc (upper limit in A2). The occurrence of
trinaries and quaternaries is analyzed only in A1. So, the content of both
independent catalogues is partly complementary and partly identical.

Our current analysis is limited to $\left\vert b\right\vert >25\deg$ because
we have verified that in the dense region $\left\vert b\right\vert \leq25\deg
$, the efficiency of selection (ratio $\beta^{A}$) based on conditions
(\ref{SA0a}) from the DR2 catalogue is even lower than in the region
$\mathbf{R}_{2}.$ And additional cuts on radial separation (similar to
Tab.\ref{Tab3} ) are not sufficiently effective. With the expected DR3 data
release, where higher accuracy of astrometric data (mainly of the parallax) is
assumed, we plan to recalculate the selection of multiple stars in the full
angular range ($4\pi$).

The content of our previous catalogue A4 has already been compared with the
A5, which contains $3,055$ binaries with magnitude $G$ $\leq13$ (\cite{AJpzkp}%
). Comparison A1 with A5 implies that only $381$ of the A5-candidates belong
to A1-region $\left\vert b\right\vert >25\deg$ and only about $60$ of which
meet selection criterion $d\leq15$ as. As shown in Tab.\ref{tab7}, similar
comparisons were made also for other catalogues. In general, the only partial
overlap of different catalogues is due to mainly different cuts in selections algorithms.

For example, the content of present catalogue A1 is compared with our previous
version A4, where the binary candidates inside the surrounding cubic region
$\left(  400\,\mathrm{pc}\right)  ^{3}$ are recorded. The candidates from the
previous catalogue that meet the conditions%
\begin{equation}
\left\vert b\right\vert >25\deg,\qquad d\leq15\,\mathrm{as} \label{SA24}%
\end{equation}
should be present in the new catalog as well. There are $25,604$ such
candidates and we succeeded $24,949$ of them to identify with binaries of the
new catalogue. So, the misidentification rate is small, $\approx2.6\%$. It can
occur, for example, with the diagram in Fig.\ref{ff17}\textit{e} that is
excluded for the new catalog, but still one pair can meet the criteria of the
previous one. At the same time, within the part of the cubic region that is
common to both catalogues, the number of binary candidates of the new
catalogue is $79,771$. It is $79,771/25,604\approx3$ times more, than the
candidates in the previous one. This ratio proves high efficiency of the
optimized method applied in the present analysis.

Our new catalogue A1 covers the region $\left\vert b\right\vert >25\deg$ up to
the distance $\lesssim3,000\,\mathrm{pc}$ and contains about $10$ times more
candidates than the previous A4, which covered full $4\pi$, but only up to
$\lesssim340\,\mathrm{pc}$. A total of almost $10^{6}$ candidates are recorded
in both merged catalogues. For more detailed comparison of the catalogues A1,
A3, A4, A5 and A6, which are based on the\textit{ Gaia} DR2 data, we have
created the merged catalogue. This catalogue is the list of pairs consisting
of four items: order number of binary, mask 103456 denoting origin from
A1,A3,A4,A5,A6 and two DR2 sources ID. The catalogues A1, A4 and the merged
one are available in the csv form on the website \url{https://www.fzu.cz/~piska/Catalogue/}.

\section{Summary and conclusion}

\label{SEC5}With the use of the optimized statistical method for analysis of
2D patterns we studied occurrence of wide multiple stars: binaries, trinaries
and quaternaries. The candidates are selected using the astrometric data
collected in the \textit{Gaia}-DR2 catalogue. So, we have studied the pairs
with angular separation wider than $\approx0.5\,\mathrm{as}$, which is present
\textit{Gaia} lower limit for resolution of two sources. Our present analysis
covers the region of galactic latitude $\left\vert b\right\vert >25\deg$ and
radial distance $L\lesssim15,000\,\mathrm{pc}$. In this space we have analyzed
about $1.3\times10^{7}$ circle events of diameter $144\,\mathrm{as}$ involving
$7.5\times10^{7}$ sources. The circle shape is advantageous for calculation of
the random background. The total number of processed sources with positive
parallax is about $1.2\times10^{8}$.

The analysis is focused on two basic parameters related to any pair of sources
in the multiple systems: angular separation $d$ and collinearity $\alpha$ of
their proper motion. Distribution of these parameters is compared with
distributions generated by the random background. The domain of the clear peak
of binaries is limited by conditions (\ref{SA0a}) that serve as the cut for
the selection of candidates. The exact knowledge of the background allows us
to define probabilistic parameter $\beta$ representing the quality of
candidates. Additional condition, which is required for the radial separation
$\Delta L$, improves quality of candidates. After this selection (enriched
exposition) the candidates are recorded in the attached catalogue. Total
numbers of candidates of wide multiple stars in the region $\left\vert
b\right\vert >25\deg$ are shown in Tab.\ref{tab5}. The catalog is compared
with some other catalogues of wide binaries selected from the \textit{Gaia} data.

We have also shown that the results of the present 2D analysis are fully
consistent with our previous 3D analysis of \textit{Gaia} DR2 data. We confirm
that the projection of wide binary orbit satisfies approximate relation
(\ref{SA15}). The average period and mass of the wide binary systems
(\ref{SA3}) are very similar to our previous estimates based on 3D analysis.

\ 

\begin{acknowledgments}
This work has made use of data from the European Space Agency (ESA) mission
\textit{Gaia} (\url{https://www.cosmos.esa.int/gaia}), processed by the
\textit{Gaia} Data Processing and Analysis Consortium (DPAC,
\url{https://www.cosmos.esa.int/web/gaia/dpac/consortium}). Funding for the
DPAC has been provided by national institutions, in particular the
institutions participating in the \textit{Gaia} Multilateral Agreement. The
work was supported by the project LTT17018 of the MEYS (Czech Republic). We
are grateful to A.Kup\v{c}o for the critical reading of the manuscript and
valuable comments. Further, we are grateful to J.Grygar for his deep interest
and qualified comments and O.Teryaev for useful discussions with interesting
ideas. We thank S.A. Sapozhnikov for providing the link to their catalogue
that we used for comparison.
\end{acknowledgments}

\end{document}